\documentclass[
    ,final            % use final for the camera ready runs
  ]
  {aipproc}

\input{aipcheck}

\layoutstyle{6x9}

\usepackage{graphicx}

\begin{document}

\newcommand{\pp}{\mathbf{p}}
\newcommand{\qq}{\mathbf{q}}
\newcommand{\mpi}{m_{\pi}}
\newcommand{\mpii}{m_{\pi^0}}
\newcommand{\beq}{\begin{equation}}
\newcommand{\eeq}{\end{equation}}

\title{QCD tests from pion reactions\\ on few-nucleon systems}

% choose from this list:
%                \texttt{http://www.aip..org/pacs/index.html}>}

\author{C. Hanhart\footnote{Talk presented by U.-G. Mei\ss ner}}{
  address={Institut f\"{u}r Kernphysik, J\"ulich Center
            for Hadron Physics, and \\ 
Institute for Advanced Simulation,
            Forschungszentrum J\"{u}lich, D--52425 J\"{u}lich, Germany}
}

\begin{abstract}
We show on two examples, namely
a calculation for charge symmetry breaking in $pn\to d \pi^0$
that allows one to extract the quark mass difference induced
part of the proton--neutron mass difference
and  a high precision calculation
for pion--deuteron scattering and its implications for the
value of the charged pion--nucleon coupling constant, how QCD tests
can be performed from low energy hadronic observables.
\end{abstract}

\maketitle

%%%%%%%%%%%%%%%%%%%%%%%%%%%%%%%%%%%%%%%%%%%%
%% MAINMATTER
%%%%%%%%%%%%%%%%%%%%%%%%%%%%%%%%%%%%%%%%%%%%

\section{Introduction}

In the last decades the effective field theory for
low energy phenomena within the Standard Model, Chiral Perturbation Theory (ChPT),
has developed to a mature tool to study hadronic
phenomena at low energies with a clear cut 
connection to QCD --- see Refs.~\cite{pipi,piN,NN} for
recent reviews with emphasis on the $\pi\pi$, the single nucleon and the
two--nucleon sector, respectively.  

\begin{center}
\begin{figure}[tb]
\includegraphics[width=145mm]{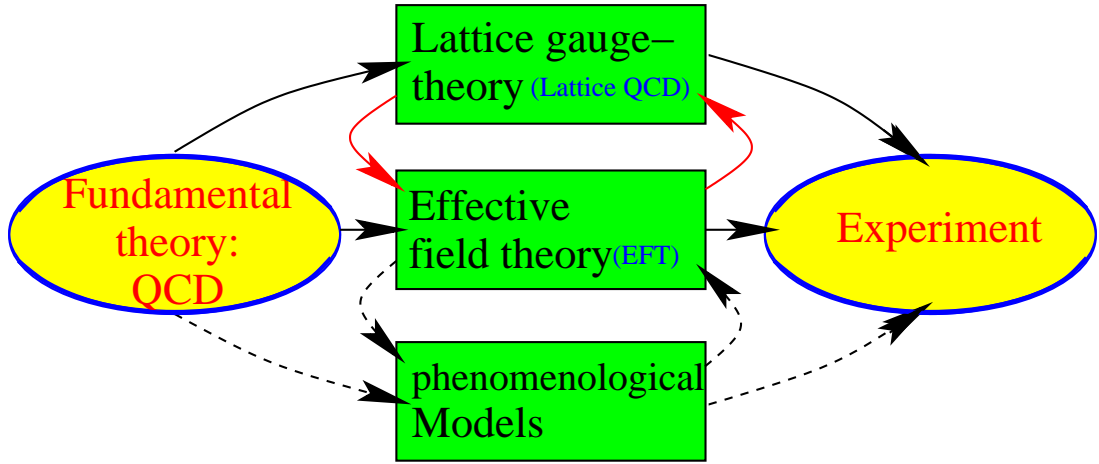}
\caption{Illustration of different methods to
compare properties of QCD to experiment.\label{fig:general}}
\end{figure}
\end{center}

One very useful application of ChPT is its use to extract from complex
reactions more fundamental quantities that can be compared to QCD predictions
straight forwardly. Those QCD predictions are calculated from first principles
using lattice gauge theory techniques~\cite{latticereview}.  Since those are
quite involved numerically the described interplay of effective field theory
and numerical methods is very rewarding.  For a long time hadronic reactions
were studied using models. Although very successful in providing a qualitative
picture of the reaction mechanisms, it is not possible to determine the
accuracy of the calculation.  Here effective field theories are in a clear
advantage: since in their very nature they are controlled expansions in some
small parameter, they allow for uncertainty estimates. This is why the role of
model calculations is decreasing in recent years. The interplay of models,
ChPT and lattice QCD is illustrated in Fig.~\ref{fig:general}.

In this presentation two examples for the strategy outlined will be described.
On the one hand a calculation will be sketched where from an analysis of
the isospin violating forward--backward asymmetry of $pn\to d\pi^+$
the quark mass induced piece of the proton--neutron mass difference was
extracted~\cite{filin}, and on the other hand a high precision calculation
is discussed that allowed for an improved  extraction of the  $\pi N$
scattering lengths~\cite{pid} from an analysis of high accuracy pionic
deuterium data.  

\section{Isospin violating $NN\to d\pi$}

Many of the isospin violating observables are dominated by effects from the
pion mass difference, since, although $m_{\pi^0}-m_{\pi^+}$ is typical for
hadronic mass differences within isospin multiplets, isospin violating effects
are enhanced due to the small pion mass. This is the reason for the importance
of charge symmetry breaking (CSB) reactions (under charge symmetry up and down
quark get interchanged), for here the pion mass difference does not
contribute and therefore effects from different quark masses get enhanced.

In this section we focus on the CSB null observable $A_{fb}(pn\to d\pi^0)$ --- the
forward--backward asymmetry in $pn\to d\pi^0$. This is a null observable since
in
a charge symmetric world the final state fixes the total isospin to 1 and
therefore the initial state gets projected on isospin 1 and is to behave as
a proton--proton/neutron--neutron pair where forward and backward are not
defined.
At TRIUMF this observable was found to be~\cite{Opper}
\begin{equation} 
\label{oppersnumber}
A_{fb}=[17.2
\pm 8 {\rm (stat.)} \pm 5.5 {\rm (sys.)}] \times 10^{-4} \ .
\end{equation} 
In this section we will briefly describe the first complete NLO
calculation for this reaction within ChPT.
Also another CSB null--observable was measured recently, namely the
total cross section for $dd\to \alpha \pi^0$~\cite{Stephenson}, however, since
the four--nucleon dynamics involved in the reaction is a lot
more complicated, no complete theoretical analysis exists 
yet for this reaction --- see Refs.~\cite{ddalpha1,ddalpha2} for some preliminary studies.

\begin{figure}[tb]
\centering
\includegraphics[width=60mm]{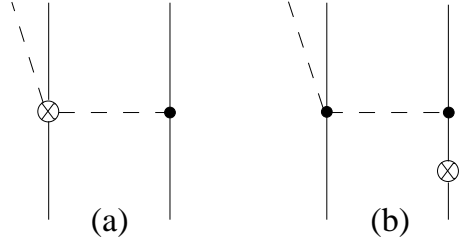}
%\includegraphics[scale=1,clip=]{diagLO_pwave.eps}
%\includegraphics[scale=0.9,clip=]{diagLO.eps}
%\vspace*{8pt}
\caption{\label{csbdiags}Leading order diagrams for the isospin
  violating $s$-wave amplitudes of $pn\to d\pi^0$.
%Isospin violating operators are denoted by $\otimes$.
 Solid (dashed) lines denote
 nucleons (pions). Diagram (a) corresponds to isospin
 violation in the $\pi N$ scattering vertex explicitly whereas  diagram (b)
 indicates an isospin-violating contribution due to the neutron--proton 
 mass difference in conjunction with the time-dependent
  Weinberg-Tomozawa operator (see text for details).  }
\end{figure}

\begin{figure}
  \includegraphics[width=.5\textwidth]{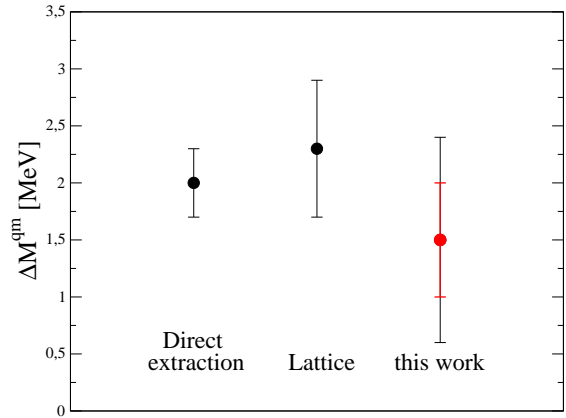}
  \caption{\label{comp} Comparison of different extractions of the
quark mass induced neutron--proton mass difference.
The points are from Refs.~\cite{Gasser} (Direct extraction), \cite{lattice}
  (Lattice), and \cite{filin} (this work). The inner (red) error
bars on the last point refer to purely the theoretical uncertainty.}
\end{figure}

The calculation described here became possible due to recent advances
in developing a systematic power counting for reactions of the type $NN\to
NN\pi$ that was complicated by the presence of the large initial momentum
$p_{\rm thr}=\sqrt{m_\pi M_N}$, with $m_\pi$ ($M_N$) for the pion (nucleon) mass, which calls for a different
expansion parameter, namely~\cite{cohenetal,threenucforce}
$$
\chi_{\rm prod}={\frac{p_{\rm thr}}{\Lambda_\chi}}=\sqrt{\frac{m_\pi}{M_N}} \ ,
$$
where for the last identity the chiral symmetry breaking scale $\Lambda_\chi$
was identified with the nucleon mass.
Nowadays the ChPT calculations have basically replaced the phenomenological
calculations~(see Ref.~\cite{polpiprod} and references therein) that dominated the field before.
For a recent review see Ref.~\cite{report}.
The reason why $A_{fb}$ is linked to the proton--neutron mass difference is
that the transformation properties of the quark mass term in QCD under axial
rotations dictates a link between mass differences of heavy hadrons and
the isospin violating pion scattering off the very same hadrons~\cite{weinberg,bira,sven}.
For the case of the $pn\to d\pi^0$ this was first studied in
Refs.~\cite{kolck,bolton}, however, these calculations were not complete,
for besides diagram (a) of Fig.~\ref{csbdiags}, where the mentioned isospin violating $\pi N$
scattering
enters, also diagram (b) enters. It gives a non--vanishing contribution since
the
isospin conserving $\pi N$ interaction is energy dependent and therefore gets
sensitive to the different energy transfer in $\pi^+$ exchange, equal to $m_\pi/2+M_p-M_n$, and in $\pi^-$ 
exchange, equal to $m_\pi/2+M_n-M_p$. In general CSB due to electromagnetism and due to quark mass
differences enter with similar strength. Here, however,
it so happens that the sum of diagram (a) and (b) are proportional only
to $\delta M^{\rm qm}$ --- the strong part of the proton--neutron mass
difference. Using the results of Ref.~\cite{pwaves} as input, the calculation revealed
 \begin{equation} 
A_{\rm fb}^{\rm LO} = (11.5 \pm 3.5)\times 10^{-4} \ 
\frac{\delta M^{\rm qm}}{\rm MeV} \ .
\label{AfbLO}
 \end{equation} 
The calculation sketched refers to a leading order calculation, however, all
contributions at NLO, namely one loop diagrams with virtual photons,
cancel~\cite{ddalpha2} --- the reason for this cancellation is now
understood~\cite{pp2dpi}.
Thus the uncertainty was estimated to be of order $\chi_{\rm prod}^2\sim 30\%$.
Using the experimental result of  Eq.~(\ref{oppersnumber}), Eq.~(\ref{AfbLO}) may be converted to give
 \begin{equation} 
\delta M^{\rm qm} = 
\left(1.5 \pm 0.8 \ {\rm (exp.)} \pm 0.5 \ {\rm (th.)}\right) \ {\rm MeV} \ ,
 \end{equation} 
where the first (second) uncertainty follows from the uncertainty of
the experiment (calculation). 
In Fig.~(\ref{comp}) this result is compared to previous extractions:
one directly from the proton--neutron mass difference using the Cottingham
sum rule to quantify the electromagnetic contribution~\cite{Gasser} and
one from lattice QCD~\cite{lattice}. Note, the calculation presented
has a comparable accuracy to the other extractions --- thus an improved measurement would
be very desirable.

\section{$\pi d$ scattering length and its implications for $g_{\pi NN}$}
%%%%%%%%%%%%%%%%%%%%%%%%%%%%%%%%%%%%%%%%%%%%
%% Sample figure:
%%
%% The option [height=...] scales the picture to the given height,
%% without it it would be printed at its nominal size
%%%%%%%%%%%%%%%%%%%%%%%%%%%%%%%%%%%%%%%%%%%%
%

The problem of $\pi d$ scattering has been studied theoretically already for many
decades using phenomenological approaches, however, nowadays the high accuracy
of modern experiments calls for improved tools for the analysis. Especially,
a consistent treatment of strong and electromagnetic few--body effects is essential for
a controlled extraction of the quite small isoscalar pion--nucleon scattering
length $a^+$, for especially electromagnetic effects might even outnumber 
its contribution to the $\pi d$ scattering length~\cite{mrr}.

There were various important advances that made a high accuracy calculation
for the $\pi d$ scattering length, reported in Ref.~\cite{ourletter},
possible: various subleading contributions were shown to
vanish~\cite{beaneetal}, there exists a calculation for $\pi N$ scattering of
the necessary accuracy~\cite{martins}, the role of various few body
corrections is understood~\cite{numericct,analyticct,susannas}, the role of the nucleon recoils is
understood~\cite{recoil,recoil_kd}, and dispersive and Delta corrections are
nowadays under control quantitatively~\cite{disp,delta}.  In this section we
will briefly sketch the results of Ref.~\cite{ourletter}, with special
emphasis on isospin violating parts.

The data for hadronic scattering lengths is best deduced from high accuracy
measurements of pionic atoms~\cite{detlev} together with properly improved
Deser formulae~\cite{LR00,mrr2} --- for a recent review see Ref.~\cite{hadatoms}.

\begin{figure}
\centering
\includegraphics[width=\linewidth]{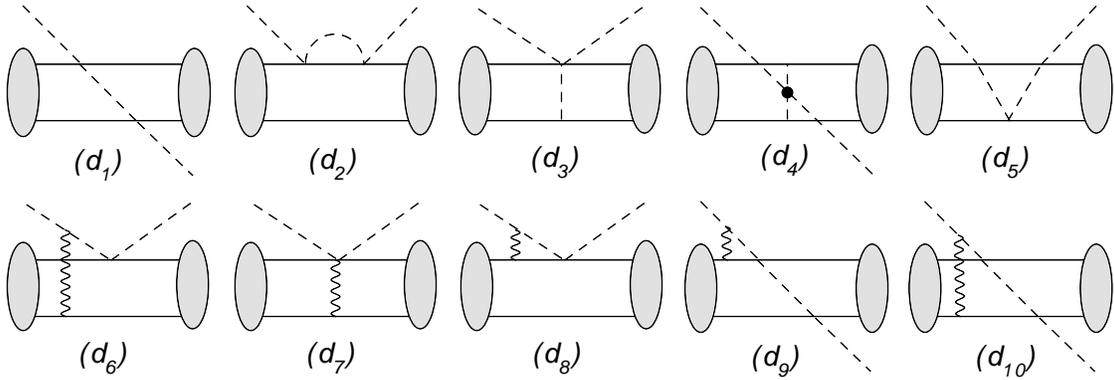}
\caption{Topologies for $\pi^- d$ scattering. 
Solid, dashed, and wiggly lines denote nucleons, pions, and photons, respectively. 
The blobs indicate  the deuteron wave functions.}
\label{fig:Feynman}
\end{figure}

The theoretical limit for the accuracy of a calculation of the $\pi NN\to \pi
NN$ transition operator is set by the first $(\bar NN)^2\pi^2$--counter term. In
the power counting of Ref.~\cite{susannas} it appears at $\mathcal{O}(\chi^2)$
relative to the leading two--nucleon operator shown in Fig.~\ref{fig:Feynman}
($d_1)$, with $\chi=m_\pi/M_N$.  We thus aim at a calculation with up--to and
including $\mathcal{O}(\chi^{3/2})$ --- square root orders appear due to the
connection between pion production (see previous section) and the dispersive
corrections~\cite{disp} as well as the numerical proximity of the
Delta-nucleon mass difference and $p_{\rm thr}$ introduced above~\cite{delta}.
The diagrams that contribute up to this order, besides those for Delta and
dispersive terms, are shown in Fig.~\ref{fig:Feynman}.  Naively, one might
expect the most important isospin violating contributions to $\pi d$
scattering from different pion masses in the leading contributions, especially
in the diagrams shown in Fig.~\ref{fig:Feynman} $(d_1)$ and $(d_2)$.  However, it is
the subtle interplay of one--nucleon and two--nucleon operators, driven by the
Pauli principle, already discussed in Ref.~\cite{recoil}, and the
orthogonality of the nuclear wave functions~\cite{recoil_kd}, that strongly
suppresses these effects. 

More difficult is the treatment of photon loops that might get enhanced due to
the masslessness of the photon together with the smallness of $\epsilon$, the deuteron
binding  energy.  For example, one finds for the leading contributions of diagrams
$(d_6)$ in an expansion in small momenta
$$
M_{d_6}\sim {-a^-}\int  \frac{d^3p\,d^3q{\Psi^\dagger(\pp-\qq)\Psi(\pp)}
}{\qq^2(\qq^2+2 m_\pi(\epsilon+\pp^2/M_p))}\sim -\frac{8\pi}{3\sqrt{2}}{\frac{a^-}{\sqrt{\mpi\epsilon}}}
\left(1+{\cal
  O}\left(\sqrt{\frac{\epsilon}{m_\pi}}\right)\right)
$$
 and $(d_8)$ of Fig.~\ref{fig:Feynman}
$$
M_{d_8}\sim %=-\left(\frac{2  \alpha  \mpi}{\pi^2}\right)
{+a^-}\int  \frac{d^3p\,d^3q{\Psi^\dagger(\pp)\Psi(\pp)}
}{\qq^2(\qq^2+2 m_\pi(\epsilon+\pp^2/M_p))}\sim +\frac{8\pi}{3\sqrt{2}}{\frac{a^-}{\sqrt{\mpi\epsilon}}}
\left(1+{\cal
  O}\left(\sqrt{\frac{\epsilon}{m_\pi}}\right)\right) \ ,
$$
where $\Psi$ denotes the deuteron wave function.
Individually the amplitudes appear to be enhanced by a factor
$\sqrt{m_\pi/\epsilon}\sim 8$ compared to the dimension analysis
estimate, however, 
 in the sum the enhanced pieces cancel. Similar cancellations can be
observed
for the other potentially infrared enhanced contributions.

\begin{figure}
  \includegraphics[width=.8\textwidth]{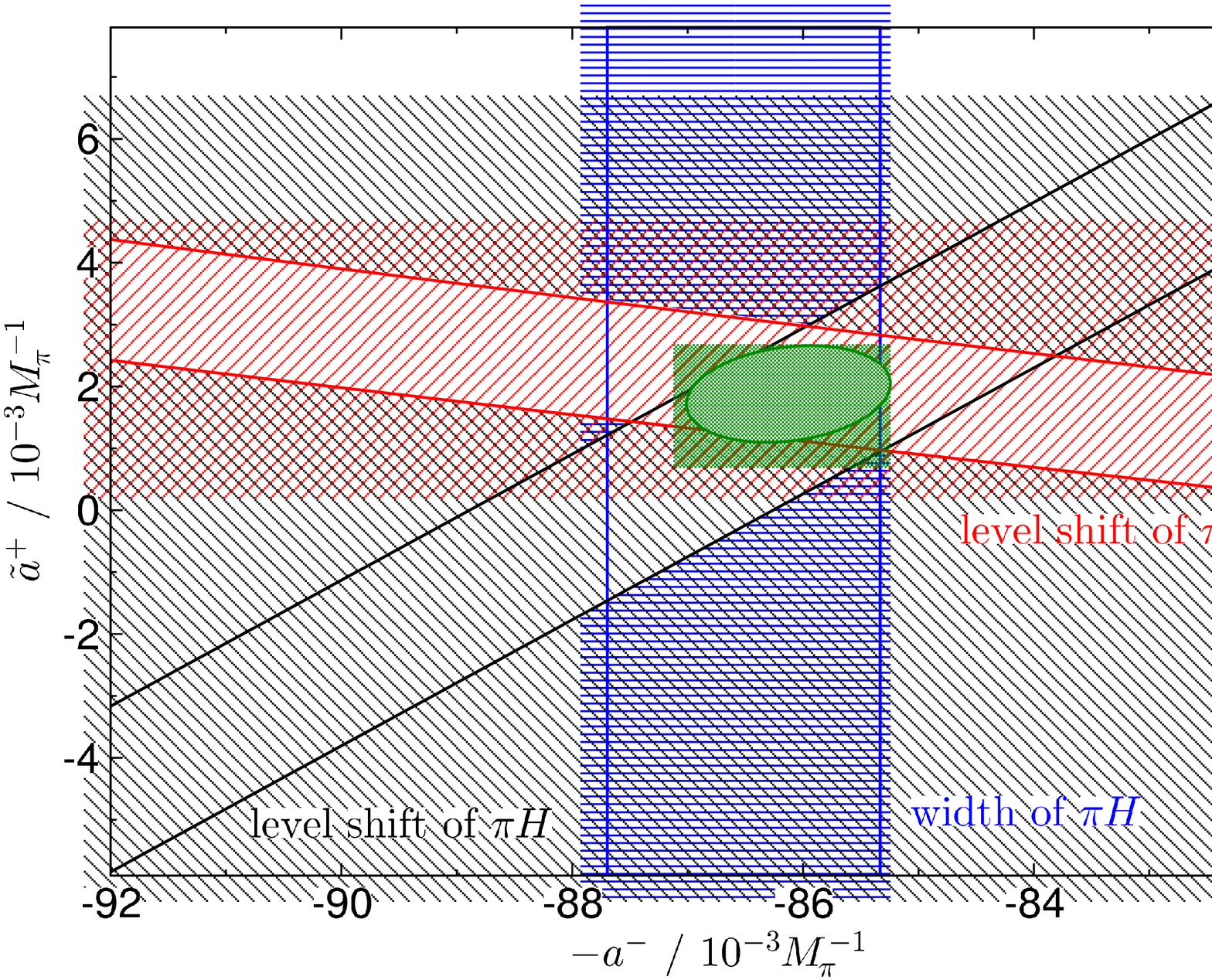}
  \caption{Combined constraints in the $\tilde a^+$--$a^-$ plane from data on the width and energy
shift of $\pi H$, as well as the $\pi D$ energy shift.\label{final}}
\end{figure}

At the end it turns out that most of the additional contributions cancel
pairwise and thus already the leading diagram --- Fig.~\ref{fig:Feynman} $(d_1)$
--- largely exhausts the value of the $\pi d$ scattering length.
The numerically most important corrections are provided by an isospin
violating
piece to the $\pi N$ scattering length and the triple scattering diagram $(d_5)$
---
from the dimensional analysis this diagram contributes at
$\mathcal{O}(\chi^2)$, it is, however, enhanced by a factor $\pi^2$
due to its special topology and thus needs to be considered~\cite{susannas}.
In addition, from pionic atoms it is not possible to extract $a^+$ directly,
but only the combination~\cite{mrr}
$$
\tilde a^+ \equiv a^+ + \frac{1}{1+\mpi/M_N}
\bigg\{\frac{\mpi^2-\mpii^2}{\pi F_\pi^2}c_1-2\alpha f_1\bigg\} \ ,
$$
with $F_\pi$ for the pion decay constant and the additional low energy constants $c_1$ and $f_1$.
The combined analysis for pionic hydrogen and pionic deuterium data yields
from the $1\sigma$ error ellipse
(c.f.  Fig.~\ref{final})
\beq
\tilde{a}^+=(1.9\pm 0.8)\cdot 10^{-3} \mpi^{-1},~~ a^-=(86.1\pm 0.9)\cdot 10^{-3}\mpi^{-1},
\label{eq:atilde+}
\eeq
with a correlation coefficient $\rho_{a^-\tilde a^+}=-0.21$.  
We find that the inclusion of the $\pi D$ energy shift reduces the uncertainty of $\tilde{a}^+$
by more than a factor of 2.
Note that in the case of the $\pi H$ level shift the width of the band is dominated
by the theoretical uncertainty in $\Delta \tilde a_{\pi^- p}$,
whereas for the $\pi H$ width the experimental error is about $50\,\%$ larger than the theoretical one.
For details on the error budget see Ref.~\cite{ourletter}.

Taken together with $c_1=(-1.0\pm 0.3)\,{\rm GeV}^{-1}$~\cite{longJOB} 
and the rough estimate $|f_1|\leq 1.4 \,{\rm GeV}^{-1}$~\cite{f1},
Eq.~\eqref{eq:atilde+} yields a non-zero $a^+$ at better than the $95\,\%$ confidence level:
\beq
a^+=(7.6\pm 3.1)\cdot 10^{-3}\mpi^{-1}.
\eeq

The final result for $a^+$ is only a little larger than several of the contributions considered in our analysis. 
This emphasizes the importance of a systematic ordering scheme, and a careful treatment of
isospin violation and three-body dynamics. 
A reduction of the theoretical uncertainty beyond that of the present analysis will be hard to achieve without 
additional QCD input that helps pin down the unknown contact-term contributions 
in both the $\pi N$ and $\pi NN$ sectors. 

As it was argued in the introduction, $\pi N$ scattering lengths are
interesting quantities by themselves, especially since they can be extracted
from lattice QCD calculations relatively easily. In the last part of this
section we will show that in addition they also provide an important link
between pion--nucleon and nucleon--nucleon scattering and in this sense a
non--trivial consistency check for our current understanding of these
fundamental reactions.  Extracted from a careful analysis of $\pi N$
scattering data, for long the charged pion nucleon coupling constant was
believed to be $g_c^2/4\pi={14.2 \pm 0.2}$~\cite{KH}.  However,
when extracted from $NN$
scattering~\cite{deSwart}, the value deduced reads $g_c^2/4\pi={13.54 \pm
0.05}$, where the error includes only the fitting uncertainty and not any
possible systematic uncertainties.
The use of the work presented to resolve the question on the value of $g_c$
becomes explicit, when using the
Goldberger-Miyazawa-Oehme (GMO) sum rule~\cite{GMO}~\footnote{In the
definition of $g_c$ there appear subtle issues due to the Coulomb
poles~\cite{david} as well as other infrared singularities. For a detailed
analysis we refer to Ref.~\cite{longJOB}.}. It reads
$$
{\frac{g_c^2}{4\pi}}{=}\left(\left(\frac{M_p+M_n}{m_\pi}\right)^{2}{-}1\right)\left[\left(1{+}\frac{m_\pi}{M_p}\right)\frac{m_\pi}{4}
\left({a_{\pi^-p}{-}a_{\pi^{+}p}}\right)
{-}\frac{m_\pi^2}{2}J^-\right] \ .
$$
Here the $\pi N$ scattering lengths, now known to higher accuracy (c.f. Fig.~\ref{final}), appear as subtraction constants for the
dispersion
integral
$$
J^-=\frac{1}{4\pi^2}\int dk\frac{{\sigma^{\rm tot}_{\pi^-p}{-}\sigma^{\rm
tot}_{\pi^{+}p}}}{\sqrt{k^2{+}m_\pi^2}} \ ,
$$
 that may be expressed in terms of observable cross sections. Values for this
integral can be taken from Refs.~\cite{ELT,AMS}. Combining the findings of
these works gives for the integral $(-1.073\pm 0.034)\,{\rm
mb}$~\cite{longJOB}, which is consistent with previous extractions (see
Ref.~\cite{arndt92}).  The GMO sum rule was used before to pin down the value
of $g_c$, however, different analyses came to different answers. While
Ref.~\cite{ELT} found a value as large as $g_c^2/4\pi=14.11 \pm 0.05 \pm
0.19$, where the first uncertainty is statistical while the second is
systematic, Ref.~\cite{arndt92} found values for $g_c^2/4\pi$ between $13$ and
$13.3$.  Also other, more general analyses from this group reported lower
values, namely $g_c^2/4\pi=13.75\pm 0.15$ \cite{arndt94}, $g_c^2/4\pi=13.76\pm
0.01$ \cite{arndt06}.  Here the uncertainties only represent the statistical
uncertainty.  No attempt was made to quantify the systematics.  More recently
Ref.~\cite{AMS} presented $g_c^2/4\pi=13.56\pm 0.36$ from a GMO analysis.
There were no further developments in the last three years and the key players
basically `agreed to disagree'~\cite{ronpriv}.  The basic improvements
provided by the analysis discussed are that for the first time isospin
violating corrections were included completely and consistently and, as the
result of using a systematic effective field theory, it became possible to
properly control the uncertainties of the $\pi N$ scattering lengths.  With
the in this way improved input we find
$${g_c^2/4\pi=13.69\pm 0.12\pm 0.15 = 13.7\pm 0.2} \ ,$$
where the first error gives the uncertainty in the scattering lengths and the
second in
the integral. From our analysis we therefore conclude that a
value for $g_c^2/4\pi$ above 14 is largely excluded.

\section{Summary}

Modern lattice QCD calculations allow for first principle calculations
of hadronic observables like scattering lengths~\cite{latticereview}.
However, in most cases those quantities are not directly accessible 
from experiments but need to be extracted from reactions with complicated
few body dynamics. In this presentation on two examples, the extraction
of the quark mass induced proton--neutron mass difference, $\Delta M^{\rm
qm}$, from the forward--backward asymmetry in $pn\to d\pi^0$, 
and the extraction of the pion--nucleon scattering lengths from
data on pionic hydrogen and deuterium, it is demonstrated that 
ChPT can be employed to extract, with controlled uncertainty, the quantities of interest from complex
reactions to allow for a comparison to lattice data
and thus for non--trivial test of QCD dynamics at low energies.

%%%%%%%%%%%%%%%%%%%%%%%%%%%%%%%%%%%%%%%%%%%%%%%%
%% BACKMATTER
%%%%%%%%%%%%%%%%%%%%%%%%%%%%%%%%%%%%%%%%%%%%%%%%

\begin{theacknowledgments}
I thank U.-G. Mei\ss ner for the presentation of the talk
and
A. Filin, F.~Ballout, V. Baru, E.~Epelbaum,
J.~Haidenbauer, M.\ Hoferichter, B.\ Kubis,
A. Kudryavtsev, V. Lensky, S. Liebig, U.-G. Mei\ss ner, A. \ Nogga,
and D.\ R.\ Phillips for the fruitful and very
educating collaboration that lead to the results presented.
The work was supported in parts by funds provided from the Helmholtz
Association (grants VH-NG-222, VH-VI-231) and by the DFG (SFB/TR 16
and DFG-RFBR grant 436 RUS 113/991/0-1) and the EU HadronPhysics2
project. 
\end{theacknowledgments}

\end{document}